\documentclass[floatfix,aps,prd,abstract,10pt,nofootinbib,preprintnumbers]{revtex4}

\usepackage{amsgen,amsmath,amstext,amsbsy,amsopn,amssymb}

\usepackage[dvips]{color}

\usepackage{graphicx}
\usepackage{epsfig}
\usepackage{multirow}
\usepackage{longtable}

\begin{document}

\newcommand{\beq}{\begin{equation}}
\newcommand{\eeq}{\end{equation}}
\newcommand{\beqa}{\begin{eqnarray}}
\newcommand{\eeqa}{\end{eqnarray}}

\title{Non-local Lagrangian bias}
\author{Ravi K. Sheth} \email{sheth@ictp.it}
\affiliation{Abdus Salam International Centre for Theoretical Physics,
             Strada Costiera 11, 34151, Trieste, Italy}
\affiliation{Center for Particle Cosmology, University of Pennsylvania, 
             209 S 33rd Street,  Philadelphia, PA 19104, USA}
\author{Kwan Chuen Chan} \email{kcc274@nyu.edu}
\author{Rom\'an Scoccimarro} \email{rs123@nyu.edu}
\affiliation{Center for Cosmology and Particle Physics, Department of Physics,  
             New York University, NY 10003, New York, USA}

\date{\today}         

\begin{abstract}
Halos are biased tracers of the dark matter distribution.  
It is often assumed that the initial patches from which halos formed 
are locally biased with respect to the initial fluctuation field, 
meaning that the halo-patch fluctuation field can be written as a 
Taylor series in the dark matter density fluctuation field.  
If quantities other than the local density influence halo formation, 
then this Lagrangian bias will generically be nonlocal; the Taylor 
series must be performed with respect to these other variables as 
well.  We illustrate the effect with Monte-Carlo simulations of a 
model in which halo formation depends on the local shear (the 
quadrupole of perturbation theory), and provide an analytic model 
which provides a good description of our results.  Our model, which 
extends the excursion set approach to walks in more than one dimension, 
works both when steps in the walk are uncorrelated, as well as when 
there are correlations between steps.  For walks with correlated steps, 
our model includes two distinct types of nonlocality:  one is due to 
the fact that the initial density profile around a patch which is 
destined to form a halo must fall sufficiently steeply around it -- 
this introduces $k$-dependence to even the linear bias factor, but 
otherwise only affects the monopole of the clustering signal.  
The other is due to the surrounding shear field; this affects the 
quadratic and higher order bias factors, and introduces an angular 
dependence to the clustering signal.  In both cases, our analysis 
shows that these nonlocal Lagrangian bias terms can be significant, 
particularly for massive halos; they must be accounted for in, e.g.,  
analyses of higher-order clustering in Lagrangian or Eulerian space.  
Comparison of our predictions with measurements of the halo bispectrum 
in simulations is encouraging.  Although we illustrate these effects 
using halos, our analysis and conclusions also apply to the other 
constituents of the cosmic web -- filaments, sheets and voids.  
\end{abstract}

\maketitle

\section{Introduction}
The virialized halos which are identified in simulations of gravitational 
clustering are biased tracers of the underlying matter field.  Typically, 
this bias is described in two ways, either by relating the halo and mass 
fields at the time the halos were identified (e.g., the present), or by 
identifying the patches in the initial conditions which are destined 
for form halos, and describing the bias between these patches and the 
initial mass fluctuation field \cite{mw1996}.  These are known as Eulerian 
and Lagrangian bias, respectively.  In either case, the simplest models 
assume that this bias is local, meaning that the biased field can be 
written as a (deterministic) function of the mass field.  However, the 
nonlinearly evolved mass field is a nonlocal function of the initial one, 
so Lagrangian and Eulerian bias cannot both be local 
\cite{clmp1998, css2012, bsdm2012}.  

It has recently been noted that neither of the two best studied 
models of Lagrangian bias, peaks theory \cite{bbks1986} and the 
excursion set approach \cite{bcek1991}, are local.  This is because 
both approaches predict that the abundance of biased tracers 
(peaks or halos) should depend, not just on the local values of 
the overdensity field, but on derivatives of the field as well \cite{ms2012}.  
This gives rise to a rather specific form for nonlocal bias, in which 
the bias is most naturally described in Fourier space, where it is 
$k$-dependent, even at the linear level \cite{dcss2010, mps2012, ps2012b}.  
The main goal of the present work is to explore models in which the 
nonlocality of bias is qualitatively different, arising only at second 
order, and associated with anisotropies in the initial field.  
This source of nonlocality is generic to models in which halos form 
from an anisotropic collapse \cite{bm1996}, for which there is 
considerable evidence \cite{smt2001}.  

Section~\ref{biasmodels} describes the relation between Lagrangian and 
Eulerian bias in nonlocal models, and Section~\ref{x6d} describes our 
excursion set based treatment of the origin of nonlocal Lagrangian 
bias, providing an analytic description of the effect.  The technical 
problem which is the subject of this section is to provide an accurate 
formula for the first crossing distribution of a barrier by 
$n$-dimensional walks with correlated steps; we use Monte-Carlo 
simulations of such walks to illustrate the accuracy of our analytic 
formulae.  Section~\ref{carmen} compares our predictions with estimates of 
the nonlocal halo bias term in numerical simulations of hierarchical 
clustering, and a final section summarizes our findings.  

\section{Nonlinear, nonlocal halo bias and its evolution}\label{biasmodels}
The excursion set model of halo abundances and evolution assumes that it is possible to identify those patches in the initial fluctuation field which are destined to form halos by a later time (e.g., the present) \cite{bcek1991}.  In what follows, we discuss why, in this case, halo bias is generally expected to be nonlocal both in the initial conditions and at later times.  

\subsection{The spherical evolution model:  Local Lagrangian bias and evolution}
The simplest implementations of this approach assume that halos form from a spherical collapse.  In this case, the initial overdensity of a patch plays an important role in determining whether or not it will form a halo \cite{gg1972}.  Almost all studies which incorporate the spherical collapse model into the excursion set approach assume that the initial overdensity is the only parameter which determines halo formation (see \cite{ms2012} for a recent exception).  As a result, these studies find that halos are locally biased versions of the initial density fluctuation field \cite{mw1996}.  If the environment of a halo is spherically symmetric, then it is natural to expect the spherical model to describe both the evolution of the patch which becomes a halo, and the evolution of its surrounding environment.  Since the spherical evolution model yields a local deterministic mapping between the initial and evolved densities, the fact that halo bias was local with respect to the initial density field means that it remains local with respect to the evolved field.  The bias factors are different of course, but they are easily calculated \cite{mjw1997}.  

Schematically, if an initial volume $V_0$ is overdense by $\delta_0$, then, in the excursion set approach, the halo overdensity (averaged over all such spheres) is 
\begin{equation}
 1 + \delta_h(m|\delta_0) \equiv \frac{\langle N_m|\delta_0\rangle}{n(m)V_0}
 = 1 + \sum_{k>0} \frac{b_k^{\rm L} \,\delta_0^k}{k!},
 \label{deltah}
\end{equation}
where the numerator is an average over all cells of volume $V_0$ and density $\delta_0$, and $n(m)$ in the denominator denotes the number density of halos of mass $m$ (i.e., the integral of the numerator over all allowed values of $\delta_0$).  The right-hand side is explicitly a function of $\delta_0$ only, so the Lagrangian bias is local.  The Lagrangian bias factors associated with the spherical model satisfy 
\begin{equation}
 \delta_c^k\, b_k^{\rm L} = \nu^{k-1}\, H_{k+1}(\nu) \qquad
 {\rm where} \quad \nu\equiv \delta_c/\sigma(m),
\end{equation}
and the $H_n$ are (the probabilist's) Hermite polynomials, and $\delta_c$ is the critical density required for spherical collapse.
The Eulerian halo overdensity is defined similarly:
\begin{equation}
 1 + \delta_h^{\rm E}(m|M,V)\equiv \frac{\langle N_m|\delta_0(M,V)\rangle}{n(m)V}
 = 1 + \sum_{k>0} \frac{b_k \,\delta^k}{k!}.
\end{equation}
The Eulerian halo bias factors are obtained upon noting that 
 $M/\bar\rho V \equiv 1+\delta$ 
and that the spherical model yields a local monotonic relation between
 $1+\delta$ and $\delta_0$, so that the expression above can be written 
as a series in $\delta$.  
Specifically, the Eulerian bias factors in the spherical collapse model are:  
\begin{eqnarray}
 b_1 &=& 1 + \frac{\nu_1^2 - 1}{\delta_1} = 1 + b_1^{\rm L}\\
 b_2 &=& \frac{8}{21}\,b_1^{\rm L} + b_2^{\rm L}\\
 b_3 &=& -\frac{796}{1323}\, b_1^{\rm L} - \frac{13}{7}\,b_2^{\rm L} + b_3^{\rm L}
 \label{localbEul}
\end{eqnarray}
etc.  Note that the Eulerian bias $b_k$ depends on the Lagrangian bias factors of equal and lower order, and the overall structure is precisely that shown as the monopole contribution to the bias in \cite{css2012}.  

Accounting for the fact that the evolution of the environment will, in general, be nonlocal means that one simply replaces the $\delta_0(\delta)$ mapping with the nonlocal one.  It is straightforward to check that carrying this through, order by order, yields the additional nonlocal bias terms given in \cite{css2012,cs2012}.  However, if halo bias is nonlocal even in Lagrangian space, then this will provide additional contributions to the Eulerian bias factors.  To see the structure of these terms, it is useful to consider models of halo formation in which factors other than the initial overdensity are important.  

\subsection{The triaxial collapse model} 
In triaxial collapse models, e.g. \cite{bm1996}, the evolution of a patch is determined by more than just its internal overdensity.  The simplest of these models uses the fact that, at each position in the initial field, one may define the deformation tensor, D, whose elements are the second derivatives of the gravitational potential.  The overdensity, which is the only quantity which matters for the spherical model, is the trace of this $3\times 3$ matrix.  So the question arises as to which (combinations) of the other elements of this matrix matter?  

To describe the shape of the gravitational potential it is common to introduce the ellipticity $e$ and prolateness $p$, defined from the eigenvalues $\lambda_i$ ($i=1,2,3$) of $\nabla_{ij}\Phi$:
\beq
 \delta_0 \equiv \lambda_1 + \lambda_2 + \lambda_3, \qquad
 e \equiv \frac{\lambda_1 - \lambda_3}{2 \delta}, \qquad {\rm and}\qquad 
 p \equiv \frac{\lambda_1 + \lambda_3 - 2\lambda_2}{2\delta}.
\label{ep}
\eeq
This set of parameters is used in triaxial evolution models of nonlinear structure formation~\cite{bm1996,smt2001}.  However, because $e$ and $p$ are ratios of the eigenvalues, it is not obvious that they are the best choice of parameters in a perturbative analysis.  In particular, one might have wondered if the rotationally invariariant quantities, 
\beqa
 I_1 &=& {\rm Tr}(D) = \sum_i \lambda_i = \delta_0, \\
 I_2 &=& \lambda_1\lambda_2 + \lambda_1\lambda_3 + \lambda_2\lambda_3, \\
 I_3 &=& {\rm Det}(D) = \prod_i \lambda_i 
\eeqa
are more relevant.  When expressed in terms of $(\delta_0,e,p)$ these are 
\beq
 I_2 = \frac{\delta_0^2}{3}[1 - (3e^2 + p^2)],\qquad
 I_3 = \frac{\delta_0^3}{27}(1-2p)[(1 + p)^2 - 9e^2]
\eeq
Since the $I_j$ do not depend on taking ratios of the eigenvalues, they, or other quantities built from them, have considerable appeal.  Indeed, ${\mathcal G}_2 = -2 I_2$ and ${\mathcal G}_3 = 6 I_3$ are the fundamental quantities in \cite{css2012}.

Another interesting combination is 
\beq
 \delta_0 = I_1, \qquad
    q_0^2 = I_1^2 - 3I_2 = \delta_0^2 (3 e^2 + p^2), \qquad {\rm and}\qquad
   u_0^3 = \frac{2I_1^3 - 9I_1I_2 + 27 I_3}{9}
       = \frac{2 \delta_0^3\, p\, (9e^2 - p^2)}{9}. \nonumber\\
 \label{dqu}
\eeq
Despite the appearance of $I_1=\delta_0$ in their definition, $q$ and $u$ are actually independent of $\delta_0$.  Moreover, they are precisely the quantities which arise in a perturbative analysis of the ellipsoidal collapse model:  $J_1$ and $J_2$ of~\cite{okt2004} are our $q_0^2$ and $9u_0^3$ respectively.  
A final combination which also arises in triaxial collapse models is 
\beq
 v = \lambda_1-\lambda_2\qquad {\rm and} \qquad w = \delta - 3\lambda_3, 
 \label{vw}
\eeq 
where $0\le v\le w$ and $0\le w\le \infty$ \cite{smlw2007}.  Like $q^2$ and $u^3$, $v$ and $w$ are also independent of $\delta$.  

Notice that $e$, $p$, $v$, $w$, $q_0^2$ and $u_0^3$ all vanish for a spherically symmetric perturbation ($\lambda_1 = \lambda_2 = \lambda_3 = \delta_0/3$).  But in general, $q_0$ follows a $\chi^2_5$ distribution \cite{st2002} (they actually called it $r$; we have changed notation to emphasize the fact that $q_0^2$ is the quadrupole which arises in perturbation theory analyses).

\subsection{Nonlocality from evolution in such models}
Whereas the spherical evolution model assumes that halos are associated with regions where $\delta_0 > \delta_c$, triaxial evolution models generically assume that $\delta_0 > \delta_c(e,p)$ or $\delta_c(q_0,u_0)$; the critical density required for collapse depends on the other elements of the deformation tensor.  Most work to date has treated the effect of $(e,p)$ approximately, by using appropriately defined average values to estimate halo abundances \cite{smt2001} and how they correlate with the large scale environment \cite{st2002}.  In this approximation, Lagrangian halo bias remains local, but the bias coefficients are modified compared to the spherical case.  In addition, halo bias at late times is also treated approximately, by assuming that the large scale environment evolves according to the spherical model.  As a result, Eulerian halo bias is also local (see \cite{sshj2001,tm2011} for explicit expressions).  If one keeps the local approximation for Lagrangian bias, but accounts for the nonlocality of nonlinear evolution of the dark matter, then this will again yield the same structure for the nonlocal terms that was described in \cite{css2012}, and shown to be robust beyond the assumption of conserved tracers. 

Before reconsidering the question of local Lagrangian bias, it is worth noting that the nonlocality of nonlinear evolution is part and parcel of the ellipsoidal collapse model.  This is because, in this model, the second and third order approximations to the nonlinear density are given by \cite{okt2004}
\beqa
 \delta^{(2)} &=& \frac{17}{21}\,\delta_0^2 + \frac{4}{21}\,q_0^2, \\
 \delta^{(3)} &=& \frac{341}{567}\,\delta_0^3 + \frac{338}{945}\,\delta_0 q_0^2 
                                         + \frac{92}{441}\,u_0^3.
 \label{d2d3}
\eeqa
Comparison with Eqs.~(34)-(36) of \cite{css2012} shows that the terms proportional to $\delta_0^2$ and $q_0^2$ in the expression for $\delta^{(2)}$ are essentially the same as their monopole $\mathcal{K}_{1,l=0}^{(2)}$ and quadrupole $\mathcal{K}_{1,l=2}^{(2)}$ (also see \cite{mr2009}).  Differences between this approximation and the exact dynamics appear for $\delta^{(3)}$, where the monopole is the same as $\mathcal{K}_{1,l=0}^{(3)}$, but the quadrupole differs. In addition, there are differences coming from nonlocal potentials that show up at third order (see e.g. Eq.~107 in \cite{css2012}), and dipole terms that correct both second and third order expressions. This shows explicitly that a self-consistent use of the triaxial evolution model should yield a better description of halo bias.  In particular, if one uses this model for the mapping between Lagrangian and Eulerian bias, then it makes sense to reconsider the assumption that the bias is local in Lagrangian space.  

\subsection{Nonlocal Lagrangian bias for triaxial collapse}
Triaxial evolution models will give rise to nonlocal Lagrangian bias if the parameters which determine $\delta_c$, e.g. $(q,u)$ or $(e,p)$, couple to the large scale environment.  To see that this is generic, consider a simple nonlocal model in which halo abundances depend on the traceless part of the initial shear field $q_0$ as well as on the initial local density $\delta_0$, and that this arises because the critical density required for collapse depends on the shear field.  This model is particularly simple because $(\delta_0/\sigma_0)$ and $(q_0/\sigma_0)^2$ are independent, the former being drawn from a Gaussian variate with unit variance and the latter from a chi-squared distribution with 5 degrees of freedom \cite{st2002}.  This simplifies the analysis required to estimate the numerator and denominator of 
\begin{equation}
 1 + \delta_h(m|\delta_0,q_0^2) \equiv
     \frac{\langle N_m|\delta_0,q_0^2\rangle}{n(m)V_0}
   = 1 + b_1^{\rm L}\, \delta_0 + b_2^{\rm L}\,\frac{\delta_0^2}{2}
       + c_2^{\rm L}\, \frac{q_0^2}{2} + \ldots
 \label{dhnonlocal}
\end{equation}
In general, notice that if we integrate over all values of $q_0^2$ at a given $\delta_0$, then this will yield 
\begin{equation}
 1 + \delta_h(m|\delta_0) \equiv
     1 + b_1^{\rm L}\, \delta_0 + b_2^{\rm L}\,\frac{\delta_0^2}{2}
       + c_2^{\rm L}\, \frac{\langle q_0^2|\delta_0\rangle}{2} + \ldots
 \label{dhnld0}
\end{equation}
Since $\langle q_0^2|\delta_0\rangle = \langle q_0^2\rangle$ does not depend on $\delta_0$, non-zero values of $c_2^{\rm L}$ will make it appear as though the $b_0$ component differs from unity.  And since $\langle q_0^2\rangle = \sigma_0^2$, this offset from unity will be vanishingly small on large scales.  

\subsection{Nonlocal Eulerian bias for triaxial collapse models}
Combining this nonlocal Lagrangian bias with Eq.~(\ref{d2d3}) for nonlocal gravitational evolution yields 
\beqa
 1 + \delta_h^{\rm E}(\delta,q^2) &=& (1+\delta) \left(1 + b_1^{\rm L}\, \delta_0
       + b_2^{\rm L}\,\frac{\delta_0^2}{2}
       + c_2^{\rm L}\, \frac{q_0^2}{2} + \ldots\right)
   = 1 + b_1^{\rm L}\, \delta_0
       + b_2^{\rm L}\,\frac{\delta_0^2}{2}
       + c_2^{\rm L}\, \frac{q_0^2}{2}
       + \delta + b_1^{\rm L}\, \delta_0\delta \nonumber\\
   &=& 1 + \delta\, (b_1^{\rm L} + 1)\,
       + \frac{\delta^2}{2} (8b_1^{\rm L}/21 + b_2^{\rm L}) 
       + \frac{q_0^2}{2} (c_2^{\rm L} - 8b_1^{\rm L}/21).
\eeqa
Since, to lowest order, $q_0^2 = q^2$, this makes the Eulerian bias factors 
\beq
 b_1 = 1 + b_1^{\rm L},\qquad
 b_2 = b_2^{\rm L} + \frac{8}{21}\,b_1^{\rm L}\qquad {\rm and}\qquad
 c_2 = c_2^{\rm L} - \frac{8}{21}\,b_1^{\rm L}.
 \label{nonlocalbEul}
\eeq
Comparison with equation~(\ref{localbEul}) shows that $b_1$ and $b_2$ are related to the Lagrangian bias factors just as in the local spherical collapse model; the nonlocality shows up as a nonzero value of $c_2$.

To make the connection to \cite{css2012} we now express this in terms of 
${\mathcal G}_2 = (2/3)(q_0^2 - \delta_0^2)$.  This makes 
\begin{equation}
 1 + \delta_h(m|\delta_0,{\mathcal G}_2) 
   = 1 + b_1^{\rm L}\, \delta_0 + (b_2^{\rm L} + (4/3)\gamma_2^{\rm L})\,
                                 \frac{\delta_0^2}{2}
       + \gamma_2^{\rm L}\,{\mathcal G}_2 + \ldots,
\end{equation}
where we have defined $\gamma_2\equiv 3c_2/4$.
Since ${\mathcal G}_2 = (2/3)(q_0^2 - \delta_0^2)$, we have that  
$\langle {\mathcal G}_2|\delta_0\rangle
 = (2/3)(\langle q_0^2\rangle - \delta_0^2) 
 = (2/3)(\sigma_0^2 - \delta_0^2) \to -(2/3)\,\delta_0^2$ on large scales.  
So, 
\begin{equation}
 1 + \delta_h(m|\delta_0) 
   = 1 + b_1^{\rm L}\, \delta_0 + b_2^{\rm L}\,\frac{\delta_0^2}{2}
       + (4\gamma_2^{\rm L}/3) \frac{\langle q_0^2\rangle}{2}\ldots.
 \label{dhnld0q0}
\end{equation}
Similarly, 
\beqa
 1 + \delta_h^{\rm E}(\delta,{\mathcal G}_2)  &=& 1 + \delta\, (b_1^{\rm L} + 1)\,
       + \frac{\delta^2}{2} (8b_1^{\rm L}/21 + b_2^{\rm L}) 
       + \frac{\delta^2 + 3/2 {\mathcal G}_2}{2} (c_2^{\rm L} - 8b_1^{\rm L}/21)
       \nonumber\\
   &=& 1 + \delta (b_1^{\rm L} + 1)\,
       + \frac{\delta^2}{2} (b_2^{\rm L} + 4\gamma_2^{\rm L}/3) 
       + {\mathcal G}_2 (\gamma_2^{\rm L} - 2b_1^{\rm L}/7),
 \label{dhnld0G2}
\end{eqnarray}
making the Eulerian bias factors 
\beq
 b_1  =  1 + b_1^{\rm L}, \qquad
 b_2  =  \frac{8}{21}\,b_1^{\rm L} + b_2^{\rm L} + \frac{4\gamma_2}{3}\qquad
 {\rm and}\qquad \gamma_2  =  \gamma_2^{\rm L} - \frac{2}{7}\,b_1^{\rm L}.
\eeq
If averaging the distribution of ${\mathcal G}_2$ at fixed $\delta$ yields the same as in Lagrangian space, i.e., 
 $\langle {\mathcal G}_2|\delta\rangle = -2\delta^2/3$ (and we set $\sigma_0\to 0$) then equation~(\ref{dhnld0G2}) implies that 
\begin{equation}
 1 + \delta_h(m|\delta) = 
     1 + b_1\, \delta + (b_2 - 4\gamma_2/3)\,\frac{\delta^2}{2} + \ldots
 \label{dhnld}
\end{equation}
which is consistent with  Eq.~(117) of \cite{css2012}.  
And, when $c_2^{\rm L}=0$, then our $\gamma_2 = -2b_1^{\rm L}/7$, which is consistent with Eq.~(118) of \cite{css2012}.  

In the next section we use the excursion set approach to estimate the numerator and denominator of Eq.~(\ref{dhnonlocal}), i.e. $\langle N_m|\delta_0,q_0^2\rangle$ and $n(m)$.  Expanding in a Taylor series yields predictions for the bias factors $b_n^{\rm L}$ and $c_n^{\rm L}$ (or $\gamma_n^{\rm L}$), and hence for the Eulerian bias factors.

\subsection{Cross-correlations}
The expressions above imply that the Lagrangian space cross-correlation between halos and mass is 
\beq
 \langle \delta_r\,[1+\delta_h(m|\delta_0,q_0^2)]\rangle
 \approx b_1^{\rm L}\,\langle\delta_r\delta_0\rangle
           + \frac{b_2^{\rm L}}{2}\langle\delta_r\delta_0^2\rangle
           + \frac{c_2^{\rm L}}{2}\langle \delta_r q_0^2\rangle
           + \frac{b_3^{\rm L}}{3!}\langle\delta_r\delta_0^3\rangle + \ldots
\eeq
where the average is over the joint distribution of $\delta_0$ and $q_0$ at one position, and of $\delta_0$ at another position a distance $r$ away.  This means that the term $\langle\delta_r q_0^2\rangle$ should be thought of as $\langle\delta_r \langle q_0^2|\delta_0\rangle\rangle$, where the inner average is over values of $q_0$ at fixed $\delta_0$, and the other average is over all $\delta_r\delta_0$ pairs separated by $r$.  This shows that replacing $\delta_h(m|\delta_0,q_0^2)$ by $\delta_h(m|\delta_0)$, its mean value for given $\delta_0$, before measuring the cross-correlation should yield the same answer as if one had included the full scatter.  This explains the agreement between no-scatter and full-scatter measurements presented in Figure~1 of \cite{cs2012}.  

For Gaussian initial conditions, all terms of the form
 $\langle\delta_r\delta_0^k\rangle$ 
in the expression above can be written as 
 $\langle\delta_0^k\langle\delta_r|\delta_0\rangle\rangle =  
  \langle\delta_0^{k+1}\rangle\, \langle\delta_r\delta_0\rangle/\langle\delta_0^2\rangle$.  
Thus, the entire local bias contribution is linearly proportional to 
 $\xi(r)\equiv \langle\delta_r\delta_0\rangle$ \cite{fs2012}.  
Of course, in the present example, $q_0^2$ and $\delta_0$ are independent, so all terms of the form $\langle \delta_r q_0^{2k}\rangle$ vanish, meaning there is no nonlocal contribution to the cross correlation.  
On the other hand, the auto correlation will receive contributions from terms of form $\langle\delta_r^{2j} q_0^{2k}\rangle$.  These will first appear at order 
$\langle\delta_r^{2} q_0^{2}\rangle$ and $\langle q_r^{2} q_0^{2}\rangle$; since they are of the same order as $\langle\delta_r^{2} \delta_0^{2}\rangle$, they will also contribute to the bispectrum.  

\section{Nonlocal Lagrangian bias in the excursion set approach}\label{x6d}
The analysis above is useful but otherwise empty formalism.  The main goal of this section is to see if the nonlocal bias factors are comparable in magnitude to the usual ones, by estimating how they depend on halo mass and time.  We use the excursion set approach to do this.  

In the excursion set approach, the abundance of objects of mass $m$ is closely related to the distribution of scales $R_m$ on which the initial overdensity $\delta > B$ for the first time, where $B\ne \delta_c$ is a measure of how the `barrier' for collapse differs from the value $\delta_c$ for the spherical evolution model.  In triaxial collapse models, $B$ is a function of $e$ and $p$ \citep{smt2001}, but, in what follows, we will use a simpler model in which $B$ is a function of $q^2$ (Eq.~\ref{dqu} shows $q^2$ is a particular combination of $e$ and $p$).  We emphasize that the logic which underlies our approach is not confined to this choice; we have chosen this simple model because it provides a particularly easy way to see the main effects associated with non-spherical collapse.  

\begin{figure}
 \centering
 \includegraphics[width=0.4\linewidth]{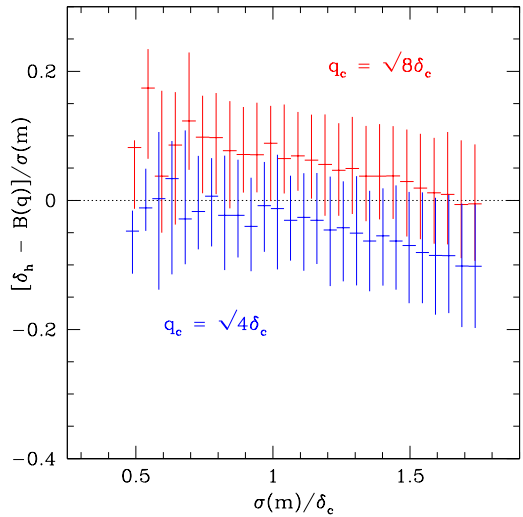}
 \caption{Difference between the actual overdensity $\delta_h$ within a protohalo in the GIF2 simulations of \cite{dts2013} and the expected overdensity given the value of the shear field (i.e. $B(q)$ of equation~\ref{Bq}), shown as a function of halo mass, for two choices of the critical value $q_c$ (smaller $q_c$ means the shear matters more).  Masses have been scaled to $\sigma(m)/\delta_c$ (large masses are on the left), and the overdensity difference has been scaled by $\sigma(m)$, as this removes most of the mass dependence of the scatter around the median relation. }
 \label{gif2Bq}
\end{figure}

The particular model we will study has
\beq
 B(q) = \delta_c\, (1 + \sqrt{q^2/q_c^2}),
 \label{Bq}
\eeq
where $q_c$ is a characteristic scale for the effects associated with the shear.  Note that non-zero $q$ always yields $B>\delta_c$, a property which will play an important role in what follows.  Equation~(\ref{Bq}) is motivated by setting $p=0$ in Eq.~(3) of \citep{smt2001}.  Figure~\ref{gif2Bq} shows that this, with $q_c^2\approx 8\delta_c^2$, provides a reasonably good (but by no means perfect) description of the actual initial (Lagrangian) overdensities in patches which collapsed to form halos by $z=0$ in the GIF2 simulations of \cite{dts2013}.  (The methods for measuring the protohalo overdensities and $q$ values are described in detail in \cite{dts2013}.)  If we use $\delta_h$ to denote the overdensity within a protohalo, then the figure actually shows $[\delta_h - B(q)]/\sigma(m)$ as a function of halo mass $m$ for two choices of $q_c$ (halo mass has been scaled to $\sigma(m)/\delta_c$, where $\sigma^2(m)$ is the variance in the initial fluctuation field with smoothed on scale $R = (3m/4\pi\bar\rho)^{1/3}$, so $\sigma(m)$ decreases as $m$ increases).  

If $\delta_h$ were equal to $B(q)$ for each halo, then $\delta_h - B(q)$ would be identically zero with no scatter.  In fact, although the mean of $\langle\delta_h-B\rangle\approx 0$, there is substantial scatter around the median value, indicating that $\delta_h$ is determined by quantities other than $q$ as well.  The scatter is larger at smaller masses, but much of this mass dependence is removed by scaling the overdensities by $\sigma(m)$.  The fact that $\langle\delta_h-B\rangle/\sigma\approx 0$ indicates that the scaling with $q$ captures much of the physics.  E.g., the mean of $(\delta_h - \delta_c)/\sigma$ is $\approx 0.2\delta_c$ (see Figure~7 in \cite{dts2013}), so including $q$-dependence does matter.  The fact that there remains a weak trend with $\sigma$ indicates that the actual scaling is not quite proportional to $q$.
But, since our main goal is to illustrate the sense of the nonlocal effects induced by effects other than the density, such as the shear, we will continue to use the simple model of Eq.~(\ref{Bq}).

The dependence of $B$ on $q$ means that we must construct a list of $\delta_i$ and $q_i$ which exhibit the correct correlations between steps, and then find the smallest $m$ (largest smoothing scale) for which $\delta_m > B(q_m)$.  (We have deliberately used $m$ as the index for the smoothing scale, since this scale is monotonically related to the inverse of the mass.)  This is straightforward to address using Monte Carlo methods \cite{bcek1991,st2002}, but, as we describe below, it turns out to be possible to write down rather accurate analytic approximations to the main results.  We will first consider the case in which the steps in the walks are uncorrelated, before describing the modifications which come from accounting for correlations.  

\subsection{Universality associated with uncorrelated steps}
While the Appendix describes a more careful analytic calculation for walks with correlated steps, we have found that the substantially simpler analysis outlined by \cite{st2002} actually provides a rather good approximation to the results for walks with uncorrelated steps, and yields considerable insight.  This analysis replaces $q^2$ with its mean value $\sigma^2$.  This means that $B(q)\to B(\sigma)$, and so the first crossing distribution associated with the six-dimensional walk (one for $\delta$ and 5 for $q^2$) should be well approximated by that for the one-dimensional condition $\delta > B(\sigma^2)$.  Reasonably accurate models for the first crossing distribution of such moving barriers are given in \cite{st2002}.  In particular, if 
\beq
 B(q) = \delta_c (1 + \sqrt{q^2/q_c^2}) \to \delta_c (1 + \sqrt{\sigma^2/q_c^2}),
 \label{Bqapprox}
\eeq
then the first crossing distribution is well-approximated by 
\beq
\nu f(\nu) \approx 2\,\left(\nu + \frac{1/4}{\sqrt{q_c^2/\delta_c^2}}\right)
 \, \frac{e^{-(\nu + \sqrt{\delta_c^2/q_c^2})^2/2}}{\sqrt{2\pi}}
 \qquad {\rm where} \qquad \nu\equiv \delta_c/\sigma .
 \label{vfv}
\eeq
In this model, the ratio $\delta_c/q_c$ is a measure of the strength of the nonlocal effects; these disappear in the limit $\delta_c/q_c\to 0$.

For similar reasons, the first crossing distribution for walks which start from some non-zero $\delta=\delta_0$ and $q^2=q_0^2$ should be well-approximated by that for one-dimensional walks which must cross 
\beq
 B(q) \to \delta_c \Bigl[1 + \sqrt{(\sigma^2 + q_0^2)/q_c^2}\Bigr] - \delta_0.
 \label{Bshifted}
\eeq
The term in the square root follows from the same logic as for the unconditioned walks; i.e., one replaces the dependence on $q^2$ by a dependence on the mean value $\langle q^2|q_0\rangle$, where the constraint $q_0$ means that $q$ is now drawn from a non-central chi-squared distribution (with 5 degrees of freedom).  This shows that the constraint enters with a plus rather than a minus sign.
%
%
%
%

Therefore, the local bias parameters are given by the usual derivatives with respect to $\delta_c$, whereas the nonlocal bias parameter is slightly more involved.  In particular, the result of using this barrier when estimating the first crossing distribution, expanding to lowest order in $\delta_0$ and $q_0^2$, dividing by the distribution associated with $(\delta_0,q_0)=(0,0)$ and subtracting one, yields:  
\beq 
 \delta_c\,b_1^{\rm L} = \nu^2 - 1 + \frac{\nu}{q_c/\delta_c}
                      + \frac{1}{1 + 4\nu q_c/\delta_c},
 \qquad 
 \delta_c^2\,b_2^{\rm L} = \nu\,H_3(\nu) + \frac{\nu^3}{q_c/\delta_c} 
     + \frac{2\nu^2}{1 + 4\nu q_c/\delta_c} - \delta_c^2\,c_2^{\rm L}
 \label{b1b2}
\eeq
and
\beq 
 \delta_c^2\,c_2^{\rm L} = -\frac{\nu^2}{(q_c/\delta_c)^2}
                        \left[1 + \nu (q_c/\delta_c) - 
                        \frac{9 (q_c/\delta_c)^2}{1 + (128/35)\nu q_c/\delta_c}\right].
 \label{c2}
\eeq 

\begin{figure}
 \centering
 \includegraphics[width=0.4\linewidth]{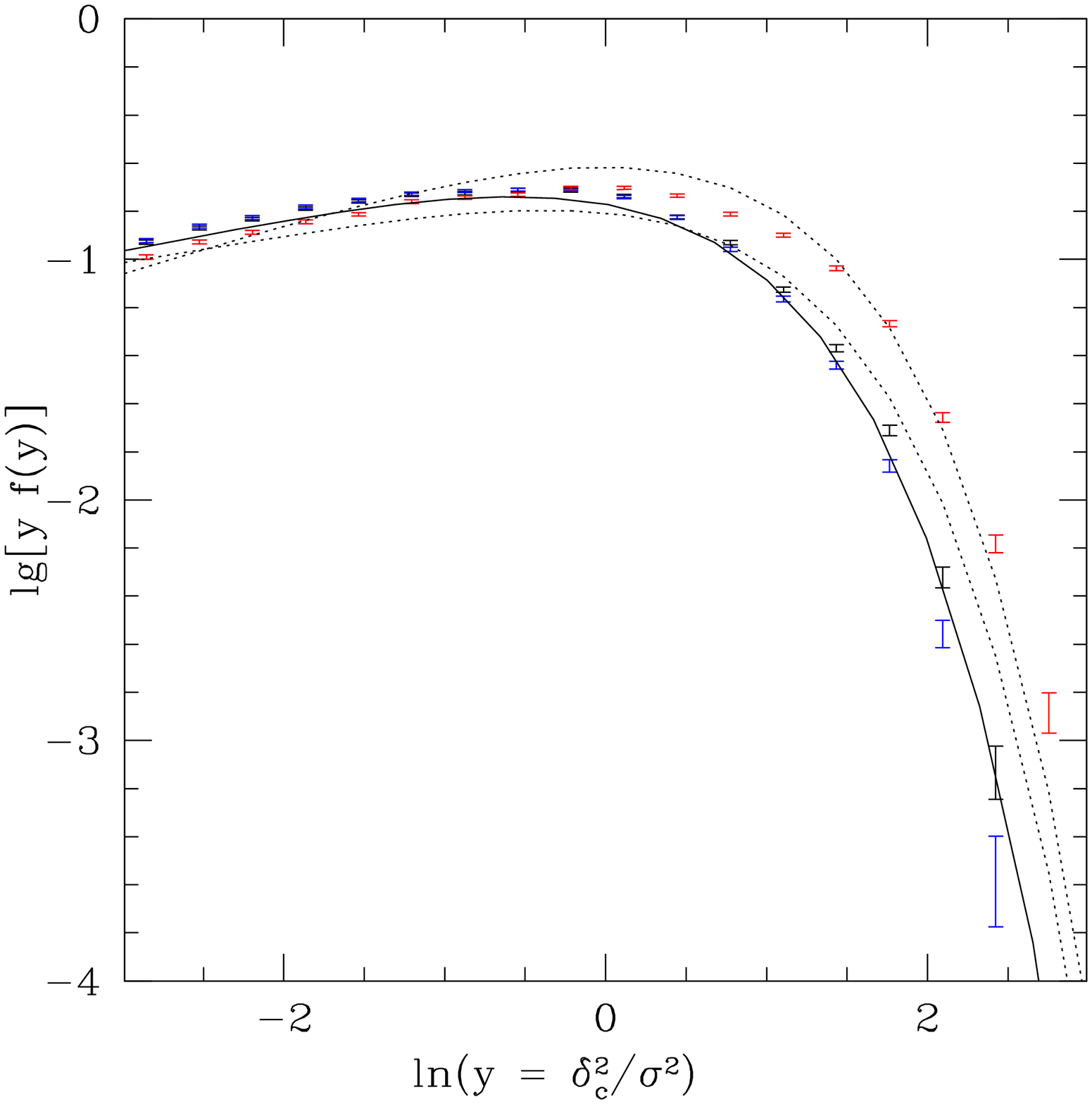}
 \includegraphics[width=0.4\linewidth]{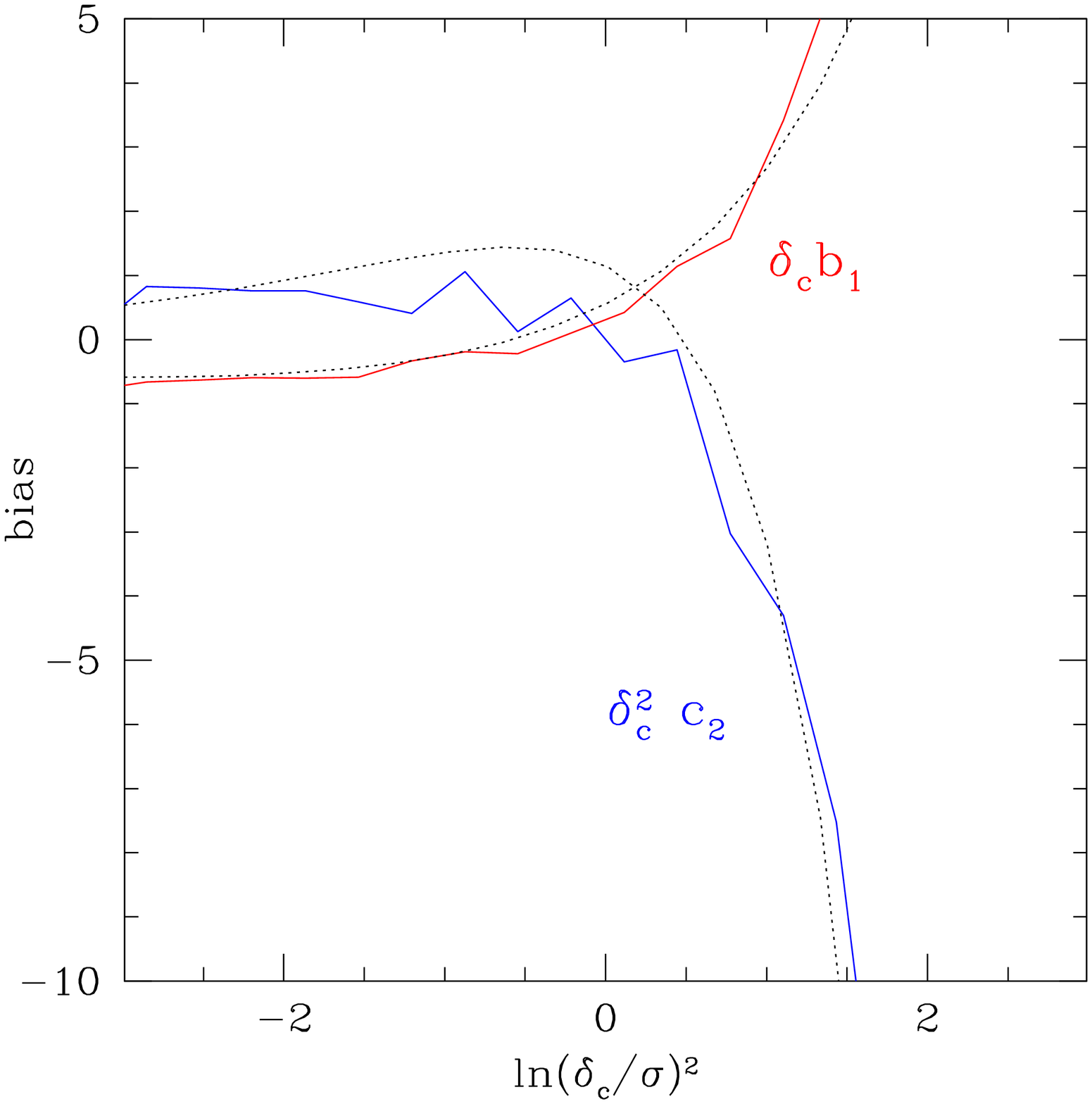}
 \caption{Excursion set description of the nonlocal Lagrangian bias which results from the local shear affecting the collapse threshold (Eq.~\ref{Bq} with $q_c^2 = 8\delta_c^2$).  Left: Distribution of first crossing scales for unconditioned walks (black); walks which began from $\delta_0/\delta_c = 0.2$ but $q_0=0$ (red); walks which began from $(q_0/q_c)^2 = 0.4$ but $\delta_0=0$ (magenta).  Dotted curves show the Press-Schechter and Sheth-Tormen distributions, and solid curve shows our Eq.~(\ref{vfv}) which follows from approximating the collapse barrier following Eq.~(\ref{Bqapprox}).  Right:  Associated large scale bias factors; dotted curves show our predictions (Eqs.~\ref{b1b2} and~\ref{c2}).}
 \label{b6rnot8}
\end{figure}

Figure~\ref{b6rnot8} shows that this works rather well.  The plot shows results for $q_c^2/\delta_c^2 = 8$, which Figure~\ref{gif2Bq} suggests is close to realistic, but we have checked that it works well for a wide range of $q_c/\delta_c$.  The panel on the left shows the distribution of first crossing scales estimated from 64000 walks:  black, red and magenta are for walks which start from 
 $(\delta_0/\delta_c,q_0^2/\delta_c^2) = (0,0)$, $(0.2,0)$ and $(0,0.4)$, 
respectively.  The first represents the unconditional distribution, whereas the second and third were chosen to isolate the effect of a large scale overdensity and (quadrupole) shear, respectively.  Notice that the model predicts more massive halos in regions with positive $\delta_0$, but fewer in regions with non-zero $q_0$.  This is a direct consequence of assuming that the barrier height is an increasing rather than decreasing function of $q^2$.  
I.e., when $\nu\gg 1$ then equation~(\ref{Bshifted}) becomes 
 $B/\sigma \to \nu \,(1 + q_0/q_c - \delta_0/\delta_c)$.
In the opposite limit, $\nu \ll (\delta_c/q_c)/(q_0/q_c)$, 
 $B/\sigma 
            \to \nu + (\delta_c/q_c) - \delta_0/\sigma$; 
for sufficiently low mass halos, only $\delta_0$ matters.  

The solid curve shows Eq.~(\ref{vfv}); it describes the unconditional distribution (i.e. black symbols) well.  The dotted curves show the Press-Schechter and Sheth-Tormen distributions; for ST, we have set $a=1$ (rather than 0.7).  Since the shape of the ST curve provides a good description of real halos, the good match between the solid and ST indicates that, for this choice of $q_c/\delta_c$, the shape of the halo mass function is like that in data.  In turn, this suggests that our model should yield realistic estimates of the sign and amplitude of nonlocal effects.  For example, this match indicates that the trend to have more massive halos in regions with positive $\delta_0$, but fewer in regions with non-zero $q_0^2$ is realistic.  

The panel on the right shows the ratio of the environment-dependent distributions to the unconditional one, scaled by $(\delta_0/\delta_c)$ and $(q_0/q_c)^2$ respectively.  The dotted curves show $\delta_c\,b_1^{\rm L}$ and $\delta_c^2\,c_2^{\rm L}$ of Eqs.~(\ref{b1b2}) and~(\ref{c2}); they describe the measurements rather well, suggesting that our analytic treatment, which accounts for the stochasticity in the barrier distribution by ignoring it following \cite{st2002}, has captured the essence of the problem.  We show elsewhere how to modify this trick for dealing with stochasticity so that it also works for other barrier shapes.  Note in particular that our analysis indicates the magnitude of $c_2^{\rm L}$ can be comparable to that of $b_1^{\rm L}$, so the Eulerian nonlocal bias parameter $c_2 = c_2^{\rm L} - 8b_1^{\rm L}/21$ can be substantial, particularly when $\nu \gg 1$ (the most massive halos).

\subsection{Departures from universality from correlations between steps}
The analysis above was based on walks with uncorrelated steps.  These correspond to smoothing the initial Gaussian field with a filter which is sharp in $k$-space.  Smoothing with filters which are more localized in real space (e.g. a spherical tophat) will result in walks with correlated steps; because such smoothing filters are intuitively closer to the physics of collapse, predictions which are based on correlated steps are expected to be more realistic (see \cite{ps2012b, smt2001, pls2012} for why this is not the full story).  

In what follows, we will show that accounting for such correlations turns out to be relatively simple, and has some important consequences.  The analysis is simplified for two reasons: $\delta$ and $q^2$ are independent whatever the smoothing filter, and the argument about transforming the six-dimensional walk problem to an effective one-dimensional barrier should work even if steps are correlated.  Recently, accurate models for the first crossing of moving barriers by walks with correlated steps have become available \cite{ms2012} (their Eq.5), so, combining these for the square-root barrier problem that is relevant here, yields 
\beq
\nu f(\nu) \approx \frac{\nu\,e^{-(\nu + \sqrt{\delta_c^2/q_c^2})^2/2}}
                            {\sqrt{2\pi}}
    \left[1 - \frac{{\rm erfc}(\Gamma\nu/\sqrt{2})}{2} + \frac{e^{-\Gamma^2\nu^2/2}}{\sqrt{2\pi}\,\Gamma\nu}\right]
 \qquad {\rm where} \qquad \nu\equiv \frac{\delta_c}{\sigma} 
 \qquad {\rm and} \qquad \Gamma^2\equiv \frac{\langle\delta'\delta\rangle^2}
                             {\langle\delta'^2\rangle\langle\delta^2\rangle},
 \label{vfvMS}
\eeq
where $\delta' \equiv d\delta/d\sigma^2$.  For $\Lambda$CDM, $\Gamma^2\approx 1/3$.  

The Appendix contrasts this with the approximation suggested in \cite{ms2012}, in which the first crossing distribution should be thought of as averaging the distribution for fixed $q$ over the distribution of $q$.  At large $\nu\gg 1$ both approximations predict a factor of 2 fewer objects than the solution for uncorrelated steps; this difference decreases as $\nu$ decreases until sufficiently small $\nu\ll (1/4)/(q_c/\delta_c)$, when the $\Gamma$-dependent factor begins to dominate.  Since $\nu\sim 0.1$ is below the regime of most cosmological interest, and it is also in the limit where the analytic approximation of \cite{ms2012} is expected to break down anyway, we will restrict attention to larger $\nu$.  In this regime the first crossing distribution will differ only slightly from that for uncorrelated steps, so we expect to find similar bias factors with the following caveats.  

First, since $\Gamma$ depends on the shape of the power spectrum, we expect plots of, e.g. $c_2$ versus $b_1$ to no longer be universal, but to depend on $P(k)$.  However, because $\Gamma$ is defined by a ratio, changes to the overall normalization of $P(k)$ will cancel out.  Second, the bias factors $b_n$ become $k$-dependent because of the correlation with the curvature term $\delta'$ \cite{ms2012}.  If $q$ were correlated with $\delta'$ then $c_2$ would also become $k$-dependent.  However, $q$ is independent of $\delta'$, and, in any case, in what follows we will restrict attention to the scales on which the $k$-dependence can be ignored.  

\begin{figure}
 \centering
 \includegraphics[width=0.4\linewidth]{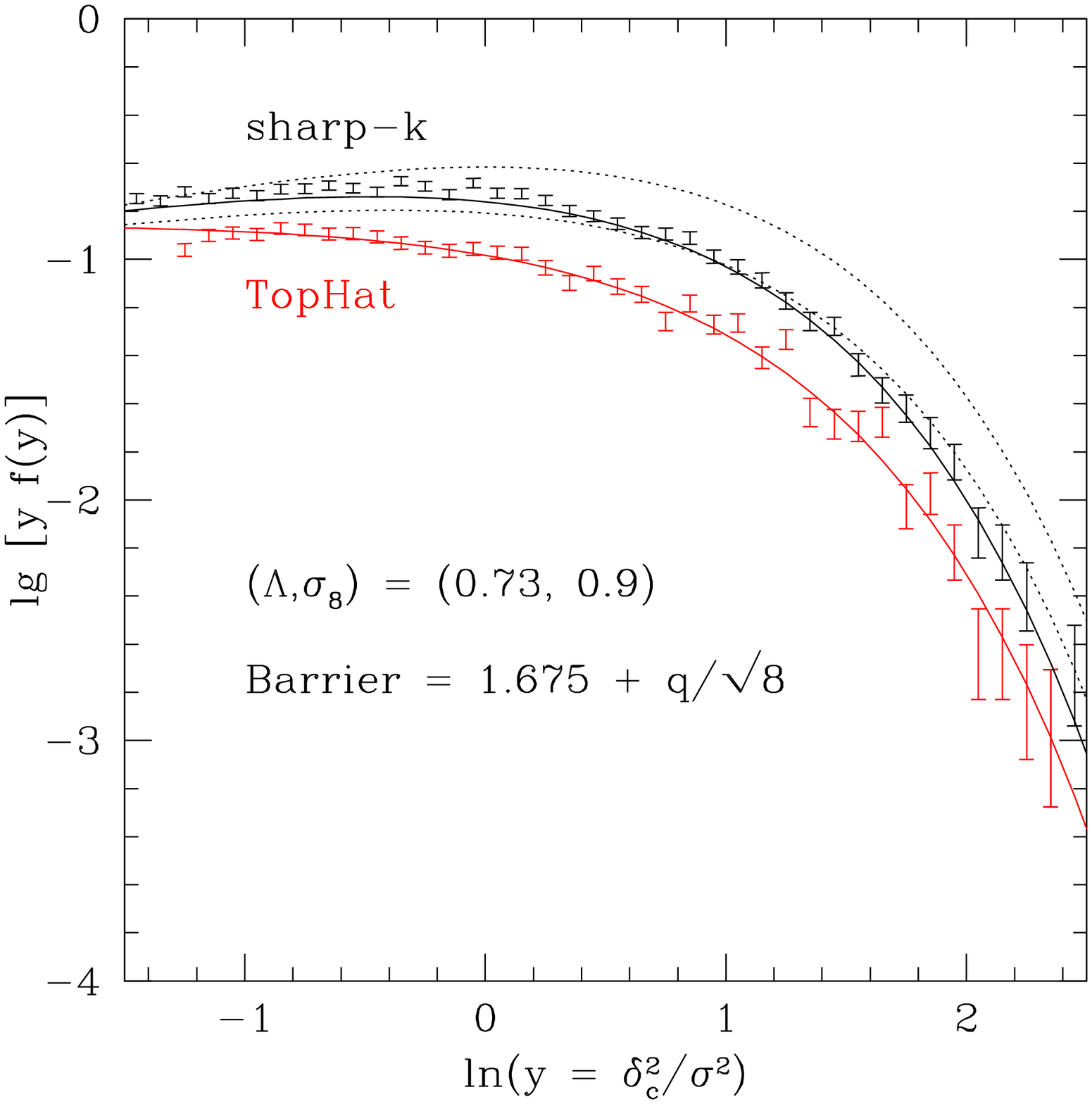}
 \includegraphics[width=0.4\linewidth]{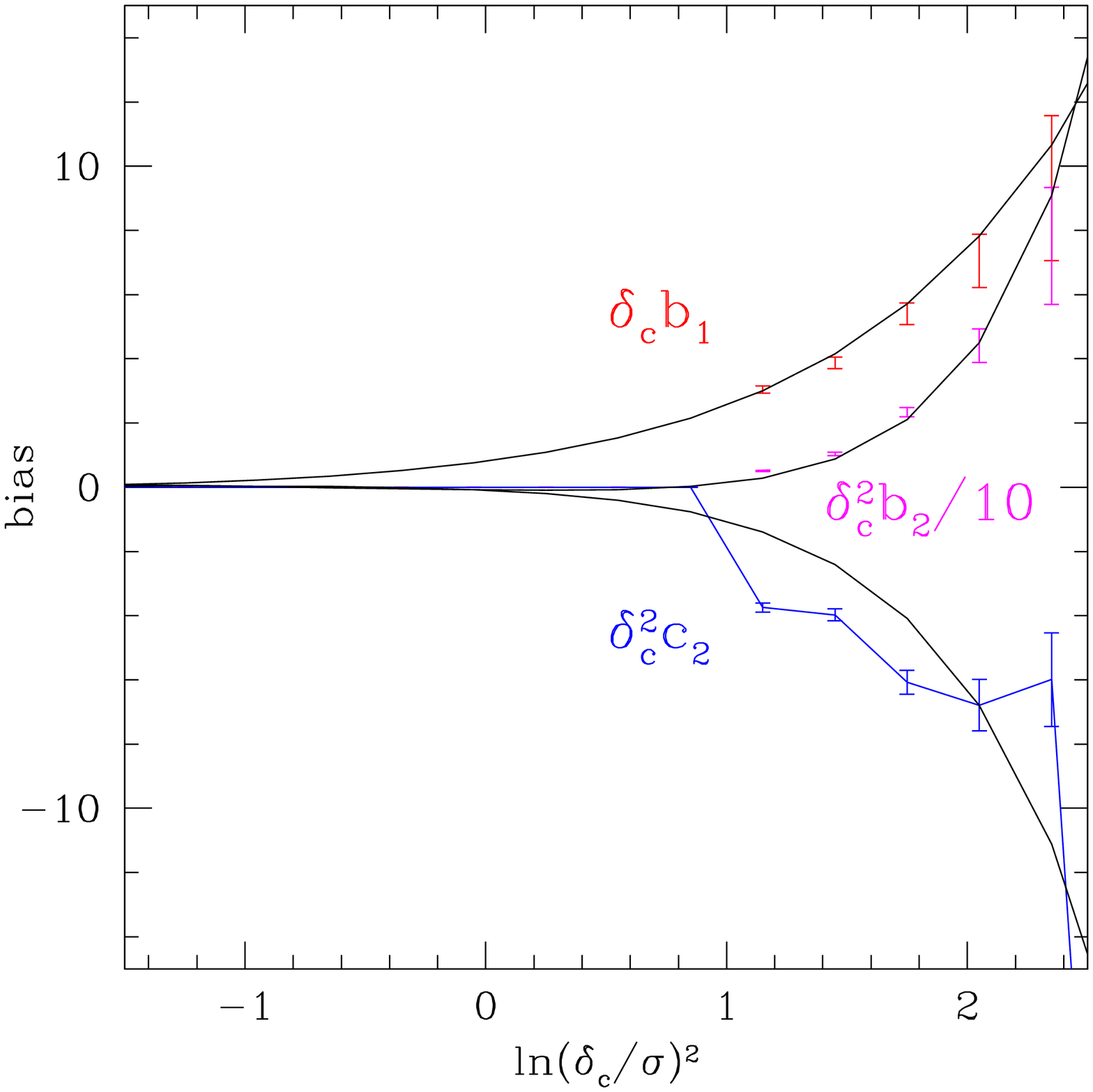}
 \caption{Comparison of excursion set predictions for the first crossing distribution (left) and associated nonlocal Lagrangian bias factors (right) for a flat $\Lambda$CDM cosmological model, when correlations between steps have been included in the analysis.  Panel on the left shows results for walks with uncorrelated steps (labeled sharp-k), and for walks in which correlations arise from TopHat smoothing.  The corresponding first crossing distributions are well-described by Eqs.~(\ref{vfv}) and~(\ref{vfvMS}), respectively (solid lines).  (The curve showing Eq.~(\ref{vfvavq}) is almost indistinguishable from that for~(\ref{vfvMS}), so we have not bothered to show it.)  We have only shown the bias factors for the TopHat smoothing filter; they are well-described by the solid lines which show Eqs.~(\ref{b1b2-corr}) and~(\ref{c2-corr}).  }
 \label{cdm6d}
\end{figure}

For walks with correlated steps that are constrained to pass through $(\delta_0,q_0)$ on scale $S_0$, the first crossing distribution can be derived similarly to when the walks started from $(0,0)$ on scale $S_0=0$.  The analog of equation~(\ref{Bshifted}) becomes 
\beq
 B(q) \to \delta_c \Bigl[1 + \sqrt{[s + (S_\times/S_0)^2(q_0^2-S_0)]/q_c^2}\Bigr]
           - (S_\times/S_0)\delta_0, \qquad
      {\rm where}\qquad \frac{S_\times}{S_0} \equiv 
      \frac{\langle\delta \delta_0\rangle}{\langle\delta_0^2\rangle}
 \label{Bcorr}
\eeq
When $\nu\gg 1$ and $S_0\to 0$ then 
 $B\to \nu [1 + (S_\times/S_0)(q_0/q_c - \delta_0/\delta_c)]$:  
except for the factor of $S_\times/S_0$, this is the same shift as for walks with uncorrelated steps (compare Eq.~\ref{Bshifted}), so we generically expect the same qualitative trends with environment.


Expanding the analog of equation~(\ref{vfvMS}) in powers of $\delta_0$ and $q_0$, and taking the $S_0\to 0$ limit yields 
\beq
 \delta_c\,b_1^{\rm L} = \nu^2 - 1 + \frac{\nu}{q_c/\delta_c} + A(\Gamma\nu),
 \qquad 
 \delta_c^2\,b_2^{\rm L} = \nu H_3(\nu) + \frac{2\nu\, H_2(\nu)}{q_c/\delta_c}
              + \frac{\nu^2}{(q_c/\delta_c)^2} + (\Gamma^2\nu^2+2)\,A(\Gamma\nu)
 \label{b1b2-corr}
\eeq
and 
\beq
 \delta_c^2\,c_2^{\rm L} = -\frac{\nu^2}{(q_c/\delta_c)^2}
          \left[1 + \nu (q_c/\delta_c) - \frac{2 (q_c/\delta_c)}{\nu}\right]
          - \frac{2\nu}{q_c/\delta_c}\,A(\Gamma\nu),
 \label{c2-corr}
\eeq
where we have defined 
\beq
 A(x) \equiv \left[1 + \frac{1 + {\rm erf}(x/\sqrt{2})}{2} 
               \frac{x\sqrt{2\pi}}{{\rm e}^{-x^2/2}}\right]^{-1}.
\eeq
The Appendix describes the large scale bias associated with equation~(\ref{vfvavq}).

Figure~\ref{cdm6d} compares these predictions with Monte-Carlos, showing that the model works rather well.  Following \cite{mps2012}, we estimate the bias factors in the Monte-Carlos using cross-correlations.  Namely, we estimate 
\beq
 b_1 = \frac{\langle sf(s|\delta_0) (\delta_0/\sqrt{S_0})\rangle}
            {\langle sf(s)\rangle\,\sqrt{S_0}\, (S_\times/S_0)},
 \qquad
 b_2 = \frac{\langle sf(s|\delta_0) (\delta_0^2/S_0 - 1)\rangle}
            {\langle sf(s)\rangle \,S_0\, (S_\times/S_0)^2},
 \qquad
 c_2 = \frac{\langle sf(s|\delta_0) (r_0^2/\langle r_0^2\rangle - 1)\rangle}
            {\langle sf(s)\rangle \,{\rm Var}(r_0^2/\langle r_0^2\rangle)\, 
             \langle r_0^2\rangle\,(S_\times/S_0)^2},
\eeq
where the angle brackets denote sums over walks which first cross at $s$, Var$(r_0^2/\langle r_0^2\rangle)$ denotes the variance of $r_0^2/\langle r_0^2\rangle$ on scale $S_0$, and the factors of $S_\times/S_0$ must be included when making estimates in this way for reasons given in \cite{mps2012}.  We present results for $S_0=0.09$, for which $S_\times/S_0\approx 1.45$.  

\begin{figure}
 \centering
 \includegraphics[ width=0.4\linewidth]{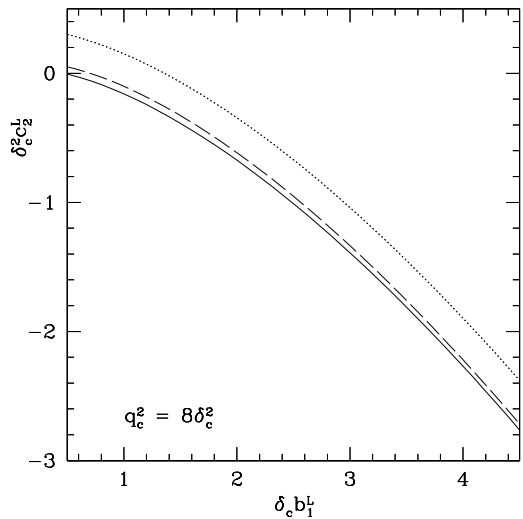}
 \caption{Dependence of excursion set prediction for the relation between $c_2^{\rm L}$ and $b_1^{\rm L}$ on the shape of the power-spectrum.  Solid and dashed lines are for correlated steps with $\Gamma^2 = 1/3$ (similar to tophat smoothing of a $\Lambda$CDM power spectrum) and 2/3, respectively; dotted line is for uncorrelated steps and is independent of $P(k)$. }
 \label{c2b1-corr}
\end{figure}

The analysis above shows that, as was the case for uncorrelated steps, $b_1$, $b_2$ and $c_2$ are all expected to depend on $\nu$.  However, there is now an additional dependence on the parameter $\Gamma$.  As a result, for walks with uncorrelated steps, there is a predicted relationship between $c_2$ and $b_1$ which is expected to be independent of the underlying cosmological parameters or power spectrum.  Correlations between steps introduce departures from this universality, so it is interesting to quantify this dependence.  Figure~\ref{c2b1-corr} shows that the predicted departures are small.  Thus, our analysis indicates that our main finding -- that the Eulerian nonlocal bias parameter $c_2 = c_2^{\rm L} - 8b_1^{\rm L}/21$ can be substantial for the most massive halos -- is robust to changes in the power spectrum.

\section{Comparison with N-body Simulations}\label{carmen}
We now compare the predicted non-locality of Lagrangian halo bias with bispectrum-based measurements of non-local bias from the Lagrangian spatial distribution of protohalos in numerical simulations of hierarchical clustering. For this purpose, we use seven Carmen realizations of the LasDamas simulations ($1120^3$ particles, in a box of size 1000 Mpc/$h$). The cosmology is a flat scale-invariant $\Lambda$CDM model with $\Omega_{\rm m}  =0.25 $, $\Omega_{\rm \Lambda}  =0.75 $, and $\sigma_8 = 0.8$.  The initial particle displacements are implemented using second order Lagrangian perturbation theory at initial redshift $z=49$. 

The halos are identified using the friends-of-friends algorithm with linking length of 0.156 of the mean interparticle separation. We consider halos with at least 20 particles and focus on halos at two redshifts, $z=0.97$ and $z=0$ respectively.  At each redshift we further divide the halos into two groups, adjusting the boundary of the mass bins so that each group has the same number density, and hence similar shot noise properties. 

To quantify non-locality of bias in Lagrangian space, we construct {\em Lagrangian} protohalos, by tracing back to the initial conditions the particles belonging to the Eulerian halos at $z=0$ and $z=0.97$.  We then use these protohalo patches to compute the cross-bispectrum with the dark matter density field at the initial conditions ($z=49$) as in \cite{css2012}. We include dark matter Fourier modes up to $k_{\rm max}^{\rm dm}=0.25 ~h$/Mpc, where the rms density fluctuation variance is 0.026 and thus tree-level perturbation theory suffices. As for the halo Fourier modes we only include up to $k_{\rm max}^{\rm h}=0.1 ~h$/Mpc, a limitation imposed by the scale where Lagrangian halo bias (as measured from the ratio of the halo-mass cross power spectrum to that of the mass) starts to show significant scale-dependence. The non-locality of bias manifests itself as an additional dependence of the bispectrum on triangle shape, which then allows us to estimate its magnitude, together with the magnitude of quadratic bias term $b_2$, and the linear bias parameter $b_1$. The latter (estimated from the cross-bispectrum) is in very good agreement with the linear bias measured from the cross power spectrum at the scales we include in the analysis.  This represents a nontrivial consistency check:  the nonlocal Lagrangian bias terms were necessary to obtain this agreement.  See~\cite{css2012} for further details of this bispectrum analysis technique.

\begin{figure}
 \centering
 \includegraphics[ width=0.5\linewidth]{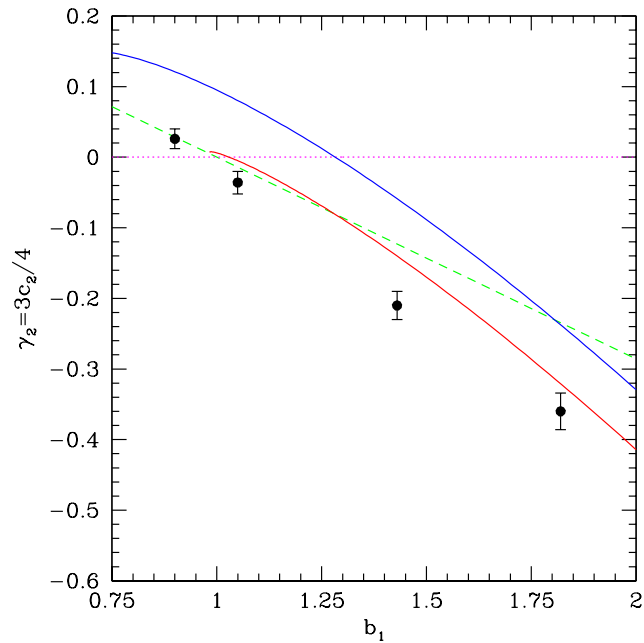}
 \caption{Comparison of the predicted relation between $c_2=c_2^{\rm L}-8b_1^{\rm L}/21$ and the Eulerian linear bias factor $b_1$ with the one estimated from bispectrum measurements of Lagrangian protohalos in N-body simulations (symbols). Solid lines show the predictions of the uncorrelated (top) and correlated (bottom) steps model; dotted and dashed lines show the local Eulerian and local Lagrangian bias models, $c_2=0$ and $c_2^{\rm L}=0$, respectively.}
 \label{gamma2resFig}
\end{figure}

Figure~\ref{gamma2resFig} presents the results, shown in terms of the Eulerian non-local bias parameter $\gamma_2=3c_2/4 = (3/4)(c_2^{\rm L} - 8b_1^{\rm L}/21)$.  Clearly, there are statistically significant deviations from the local Eulerian bias model (horizontal dotted line).  At high masses, there are significant deviations from the local Lagrangian bias model (dashed line) as well.  Upper and lower solid curves show the excursion set predictions for walks with uncorrelated and correlated steps (we set $q_c = \sqrt{8}\delta_c$).  The correlated steps prediction in particular reproduces the measured trends reasonably well.  Setting $q_c = \sqrt{4}\delta_c$ rather than our fiducial value of $\sqrt{8}\delta_c$ (see Figure~\ref{gif2Bq} for why this is also acceptable) yields slightly better agreement.  This supports the view that the model developed in the previous sections, despite its simplicity, is a fairly accurate estimate of the physics which leads to nonlocal bias effects.  

A more detailed comparison with simulations will be presented elsewhere.  E.g., one might imagine leaving $\delta_c$ and $q_c$ to be free parameters when fitting Eqs.~(\ref{vfv}) and/or~(\ref{vfvMS}) to the halo abundances in simulations, and then using these best-fit values to predict the measurements shown in Figure~\ref{gamma2resFig}.  

\section{Discussion}
If halo bias is nonlocal in Lagrangian space (Eq.~\ref{dhnonlocal}), then this will add nonlocality in Eulerian space bias (Eq.~\ref{nonlocalbEul}).  We provided an explicit calculation of this effect, in which halo formation depends on the initial local density $\delta_0$ and shear field $q_0^2$.  

In the excursion set approach, the problem of estimating halo abundances reduces to solving for the first crossing distribution of a suitably chosen barrier (Eq.~\ref{Bq}) by $6$-dimensional random walks.  In particular, we argued that a barrier which is linear in $q_0$ should provide a good first approximation to the physics of halo collapse (Figure~\ref{gif2Bq}).  And we provided a simple but accurate analytic approximation for the first crossing distribution for the case in which walks have uncorrelated steps (Eq.~\ref{vfv}); the approximation follows from treating the full 6-dimensional problem as an effective one-dimensional one (Eq.~\ref{Bqapprox}).  

Predictions for halo bias come from studying walks which do not start from the origin.  We argued that the associated first crossing distribution is best thought of in terms of a shifted barrier (Eq.~\ref{Bshifted}), from which it is straightforward to derive formulae for halo bias formulae (Eqs.~\ref{b1b2} and~\ref{c2}).  These formulae, which quantify how the large scale density and shear fields affect halo abundances as a function of mass and time, are quite accurate (Fig.~\ref{b6rnot8}).  

For walks with correlated steps, the predicted first crossing distributions and bias formulae can be written in units in which they are universal (independent of power spectrum and cosmology).  We argued that this universality should be weakly broken if steps are correlated (Fig.~\ref{c2b1-corr}).  In this case, too, we provided analytic approximations for the unconditional first crossing distribution (Eq.~\ref{vfvMS}) and halo bias factors (Eqs.~\ref{b1b2-corr} and~\ref{c2-corr}), which were quite accurate (Fig.~\ref{cdm6d}).  Our results indicate that nonlocal bias effects, as quantified by the parameter $c_2 = c_2^{\rm L} - 8b_1^{\rm L}/21$ of Eq.~(\ref{nonlocalbEul}), can be substantial for the most massive halos.

Our analysis is easily extended to describe the more complicated case in which the barrier $\delta_c$ depends on $(e,p)$ of Eq.~(\ref{ep}) rather than simply $q^2$.  This is because $\delta_c(e,p)$ is a function of the combination $e\delta$ and $p\delta$ \cite{smt2001}, and, like $q^2$, these combinations are actually independent of $\delta$ \cite{st2002}.  Alternative parametrizations of this \cite{smlw2007} have $\delta_c(v,w)$ where $v$ and $w$ (defined in Eq.~\ref{vw}) are also independent of $\delta$.  This means that the analog of Eq.~(\ref{vfvavq}), in which one averages over a distribution of first crossing distributions, remains a good approximation.  

In addition, the idea that one can map the 6-dimensional walk problem to a 1-dimensional moving barrier problem should continue to hold when $\delta_c(e,p)$ or $\delta_c(v,w)$ rather than $\delta_c(q)$, so the analog of Eq.~(\ref{vfvMS}) should also provide a reasonable approximation.  Therefore, we believe our analysis should be applicable to other parametrizations or models of the effects of nonlocality.  Indeed, our analysis should also apply to cases where one places conditions on the eigenvalues of the deformation tensor (e.g. all three have the same sign), rather than the combinations $e,p$, or $v,w$.  In this respect, it provides the basis for modelling not just halos, but the abundance and spatial distribution of superclusters, filaments, sheets and voids as well.  In particular, our analysis predicts that all of these constituents of the cosmic web should exhibit nonlocal bias effects; it will be interesting to see if such effects are discovered in simulations.  

In our analysis of nonlocal halo bias, we assumed that the effect of the large scale environment was to affect the distribution of $\delta$ and $q$ on smaller scales, but not the shape of the collapse barrier.  If the halo formation process depends on the surrounding environment, as the analysis in \cite{vd2008} suggests, then this will provide an additional contribution to nonlocal bias.  It is straightforward to include such an effect in our treatment of conditional $n$-dimensional walks, but our current expressions do not do so.  

We stated at the start that the nonlocal effects which are the subject of this paper, and which enter at the quadratic level ($q^2$, $e\delta$, etc.), are qualitatively different from those which enter even at the linear level, and contribute $k$-dependent terms to the bias.  In principle, both effects are present in our correlated walks calculation -- the latter arise from the dependence of the first crossing distribution on the derivative of the initial density field \cite{ms2012, mps2012}, rather than on the anisotropic distribution of the mass.  Since our main goal was to illustrate the effects of $q^2$, etc., we ignored these other terms, but a more complete model would include them.  Separating these terms from one another should be possible, since the terms which depend on derivatives of the field will only contribute to the monopole of the bias.  This is the subject of work in progress.  

A first comparison with numerical simulations (Figure~\ref{gamma2resFig}) showed that our model is fairly accurate at describing the magnitude of nonlocal Lagrangian bias, in particular after accounting for correlations between steps. A more detailed comparison with simulations may require merging the formalism here with the additional requirement that one is interested in special positions (such as peaks), rather than random positions in the field.  As noted in \citep{ps2012b}, the formalism of \citep{ms2012} on which our analysis is based allows the inclusion of this requirement with no additional conceptual complications; for peaks in a Gaussian field, this is particularly straightforward, because the distribution of the shear field around such peaks is known \cite{vdWB1996}.  Our analysis also suggests that a fruitful extension of the peaks model to smaller masses than that on which it usually breaks down is to look for peaks in the field defined by $\delta - \sqrt{q^2/q_c^2}$.

\section*{Acknowlegements} 
RKS is supported in part by NSF-AST 0908241 and NASA NNX11A125G.  He is grateful to G. Tormen for providing the measurements of protohalo overdensities and shear values in 2007, which were used to make Figure~\ref{gif2Bq}, and to F. Bernardeau for discussions about this project, and the IPhT group at CEA Saclay for their hospitality, during June 2010.  KCC acknowledges the support of a James Arthur Graduate Assistantship and a Mark Leslie Graduate Assistantship.  RS is partially supported by NSF AST-1109432 and NASA NNA10A171G, and thanks RKS and UPenn for hospitality during a sabbatical stay in the Fall of 2009, where much of this work was performed.  RS and RKS thank the Aspen Center for Physics, where this project began in 2007.  The simulations presented here are part of the LasDamas collaboration suite\footnote{\tt http://lss.phy.vanderbilt.edu/lasdamas}  and were run thanks to a Teragrid allocation and the use of RPI and NYU computing resources.

\appendix
\section{Analytic estimate of nonlocal Lagrangian bias}
For barriers of the form $\delta_c + q\,\delta_c/q_c$, Eq.~(13) of \cite{ms2012} yields a simple estimate of the first crossing distribution:  
\beq
 \nu f(\nu) \approx \left[1 - \frac{{\rm erfc}(\Gamma\nu/\sqrt{2})}{2} + \frac{\nu\,e^{-\Gamma^2\nu^2/2}}{\sqrt{2\pi}\,\Gamma\nu}\right] \int d\rho\,p_5(\rho)\,
  \frac{e^{-(\nu + \sqrt{\rho^2 \delta_c^2/q_c^2})^2/2}}{\sqrt{2\pi}}\,,
\eeq
where
\beq
 p_5(\rho)\,d\rho \equiv \frac{d\rho^2}{\rho^2}\,
     \left(\frac{5\rho^2}{2}\right)^{5/2}\frac{e^{-5\rho^2/2}}{\Gamma(5/2)},
 \qquad{\rm with}\qquad \rho\equiv q/\sigma.
\eeq
The integral can be written in terms of the parabolic cylinder function 
$U(a,z)$ with $a=9/2$ and $z = (1 + 5 q_c^2/\delta_c^2)^{-1/2}$, which, in 
turn, can be written in terms of derivatives of
 $e^{z^2/2}{\rm erfc}(z/\sqrt{2})$.   The integral equals 
\beq
 2\frac{e^{-\nu^2/2}}{\sqrt{2\pi}} 
  \left(\frac{5/2}{\delta_c^2/q_c^2 + 5}\right)^{5/2}
  \frac{4!}{\Gamma(5/2)}\,e^{z^2/4}\, U(9/2,z)
 = \frac{e^{-\nu^2/2}}{\sqrt{2\pi}}\, \frac{\Gamma(1/2)}{\Gamma(5/2)}
   \left(\frac{5/2}{\delta_c^2/q_c^2 + 5}\right)^{5/2}
   \frac{d^4 [e^{z^2/2}{\rm erfc}(z/\sqrt{2})]}{dz^4}
\eeq
where $z = (1 + 5 q_c^2/\delta_c^2)^{-1/2}$.
Thus, 
\beqa
 \nu f(\nu) 
 &=& \frac{\nu\,e^{-\nu^2/2}}{\sqrt{2\pi}}\,
     \left[1 - \frac{{\rm erfc}(\Gamma\nu/\sqrt{2})}{2} + \frac{e^{-\Gamma^2\nu^2/2}}{\sqrt{2\pi}\,\Gamma\nu}\right]\, \frac{(5/2)^{5/2}}{\Gamma(5/2)}
     \,\left[C_1(\nu) - C_2(\nu)\right]\,\qquad{\rm where}\nonumber\\
 C_1 &=&  \left(\frac{(\nu\,\delta_c/q_c)^4}{(5 + (\delta_c/q_c)^{2})^4}
       + \frac{6\,(\nu\delta_c/q_c)^2}{(5+(\delta_c/q_c)^{2})^3} 
       + \frac{3}{(5 + (\delta_c/q_c)^{2})^2}\right)\sqrt{\frac{2\pi}{5+(\delta_c/q_c)^{2}}}
        {\rm erfc}(y)e^{y^2/2} \nonumber\\
 C_2 &=& 5\frac{2\,(\nu\delta_c/q_c)}{(5 + (\delta_c/q_c)^{2})^3}
         + \frac{2(\nu\delta_c/q_c)^3}{(5 + (\delta_c/q_c)^{2})^4}.
\label{vfvavq}
\eeqa
Although this expression is not particularly illuminating, we have included it because the peak background split bias factors are easily estimated from the fact that
\beq
 \frac{d^m\,[e^{z^2/4} U(a,z)]}{dz^m} = (-1)^m (1/2+a)_m\, e^{z^2/4}\,U(a+m,z).  
\eeq

\begin{figure}
 \centering
 \includegraphics[ width=0.4\linewidth]{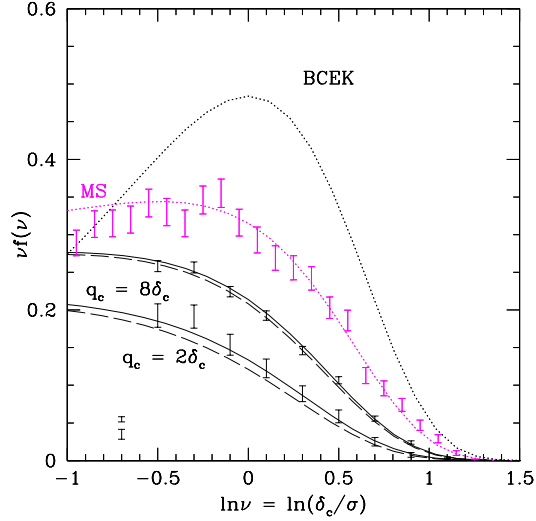}
 \caption{Dependence of excursion set prediction for the first crossing distribution of the barrier $\delta_c\,(1 + q/q_c)$ by six-dimensional walks, on $q_c$.  Solid curves show Eq.~(\ref{vfvavq}), and dashed curves show the approximation Eq.~(\ref{vfvMS}).  Curve labelled MS shows the $q_c\to\infty$ limit, and dotted curve labelled BCEK shows this limit for one-dimensional walks with uncorrelated steps.  For the walks with correlated steps, the underlying power-spectrum is $\Lambda$CDM, and the correlations are due to smoothing with a real space Tophat.  }
 \label{nonlocalMF}
\end{figure}

More insight comes from approximating the integral over $\rho$ in Eq.~(\ref{vfvavq}) with the value of the integrand at its mean value $\langle\rho\rangle = 1$.  This yields Eq.~(\ref{vfvMS}) in the main text.  A little algebra shows that this particularly simple and intuitive approximation should be quite accurate when $q_c/\delta_c\gg 1$.  Figure~\ref{nonlocalMF} shows that this is indeed the case, although equation~(\ref{vfvavq}) is accurate over a wider range of $q_c/\delta_c$.  

To estimate the large scale bias, for comparison with the peak background split estimate, we must first estimate the first crossing distribution for walks which pass through $(\Delta,Q)$ on some $S\le s$.  Then the Lagrangian bias factors are given by expanding 
\beq
 1 + \delta_h(m|\Delta,Q) \equiv \frac{f(m|\Delta,Q)}{f(m)} 
\eeq
in powers of $\Delta$ and $Q$.  
This shows that one generally expects halo abundances and hence Lagrangian bias to depend on $Q$ as well as $\Delta$.  

The effect of the constraint is to modify the distribution of $(\delta,q)$ but not the barrier (this is actually an assumption -- in principle, the shape of the barrier could depend on the large scale environment, in which case this would provide an additional environmental effect), so the calculation is actually rather similar to that for unconstrained walks.  The independence of $q$ and $\delta$ on all scales means that the conditional distribution, given the values $\Delta,Q$ on some other scale (in what follows, we will assume this other scale is larger), factorizes:
\beq
 p(\delta',\delta,q|\Delta,Q) = p_5(q|Q)\, g(\delta'|\delta,\Delta)\,
                                           g(\delta|\Delta),
\eeq
where 
\beq
 g(\delta|\Delta) = \frac{\exp^{-(\delta - \gamma\sqrt{s/S}\Delta)^2/2s(1-\gamma^2)}}
                         {\sqrt{2\pi s(1-\gamma^2)}}
   \qquad {\rm with}\qquad \gamma^2
          = \frac{\langle\delta\Delta\rangle^2}
                 {\langle\delta^2\rangle\langle\Delta^2\rangle}
          = \frac{\langle\delta\Delta\rangle^2}{sS}
\eeq
and $p_5(r|R)$ is a non-central chi-squared distribution:
\beq
 p_5(q|Q)\,dq = d\rho_\gamma\,p_5(\rho_\gamma)\,
                e^{-\frac{5\gamma^2Q^2}{2S(1-\gamma^2)}}
  \left[1 + \sum_{j=1}^\infty 
  \left(\frac{5\gamma^2 (Q^2/S)}{2(1-\gamma^2)}\frac{5\rho_\gamma^2}{2}\right)^{j}
              \frac{\Gamma(5/2)}{\Gamma(5/2 + j)}\right]
  \qquad {\rm where}\qquad \rho_\gamma^2 \equiv \frac{q^2/\sigma^2}{1-\gamma^2}.
\eeq
Notice that this is simply a $\chi^2_5$ distribution for $\rho_\gamma$ when $Q=0$.  In general, the integral over $\rho_\gamma$ can again be written in terms of parabolic cylinder functions.  But, to lowest order in $Q$,
\beq
 p_5(q|Q)\,dq = d\rho_\gamma\,p_5(\rho_\gamma)\,
       \left(1 - (1-\rho_\gamma^2)\frac{5 (\gamma^2 Q^2/S)}{2(1-\gamma^2)}\right)
  \qquad {\rm where}\qquad \rho_\gamma^2 \equiv \frac{q^2/\sigma^2}{1-\gamma^2}
\eeq
This is most easily understood by thinking of the non-central chi-squared distribution as a Poisson mixture of central chi-squared distributions of ever higher order, and keeping only the lowest two terms.  Note that, for a sharp $k$-filter, 
 $\gamma^2 = \langle\Delta^2\rangle/\langle\delta^2\rangle = S/s$, 
making
 $\gamma^2 R^2/S = R^2/s$ in the expression above.  However, we are interested in the general case, for which it is convenient to define
 $S_\times \equiv \langle\delta\Delta\rangle$ making
 $\gamma^2 R^2/S = (S_\times/S)^2 (R^2/s)$.  

\beqa
 p_5(q|Q)\,dq &=&
 d\rho_\gamma\,p_5(\rho_\gamma)\,
                e^{-\frac{5\gamma^2Q^2}{2S(1-\gamma^2)}}
  \left[\sum_{j=0}^\infty 
    \left(\frac{5\gamma^2 (Q^2/S)}{2(1-\gamma^2)}\right)^j
     \frac{\Gamma(5/2)}{\Gamma(5/2 + j)}\sum_{i=0}^j {j\choose i}\, x^i y^{j-i}\right] \nonumber\\
 &=& d\rho_\gamma\,p_5(\rho_\gamma)\,
                e^{-\frac{5\gamma^2Q^2}{2S(1-\gamma^2)}}
  \left[\sum_{i=0}^\infty \left(\frac{5\gamma^2 (Q^2/S) x}{2(1-\gamma^2)}\right)^{i}
     \sum_{j\ge i}^\infty \left(\frac{5\gamma^2 (Q^2/S)y}{2(1-\gamma^2)}\right)^{j-i}
     \frac{\Gamma(5/2)}{\Gamma(5/2 + j)} {j!\choose (j-i)! i!}\right]
\eeqa


If we define 
\beq
 \nu_{\Delta} = \frac{\delta - \gamma\sqrt{s/S}\Delta}{\sqrt{2s(1-\gamma^2)}}
   \to \frac{\delta - \gamma\sqrt{s/S}\Delta}{\sqrt{2s}}
\eeq
on large scales.  Then 
\beq
 \nu f(\nu|\Delta,Q) \approx
\nu_{\Delta}\left[1 - \frac{{\rm erfc}(\Gamma\nu_\Delta/\sqrt{2})}{2} + \frac{e^{-\Gamma^2\nu_\Delta^2/2}}{\sqrt{2\pi}\,\Gamma\nu_\Delta}\right] \int dq\,p_5(q|Q)\, \frac{e^{-(\nu_\Delta + \sqrt{q^2 \delta_c^2/q_c^2})^2/2}}{\sqrt{2\pi}}
\eeq
The purely local contributions to the bias come from considering the ratio 
$f(\nu|\Delta,Q=0)/f(\nu)$ in the $S\to 0$ limit.  These correspond to replacing $\nu\to\nu_\Delta$ in equation~(\ref{vfvavq}).  In addition, the expression above shows that if the barrier does not depend on $q$, then $ \nu f(\nu|\Delta,Q)$ does not depend on $Q$; nonlocal effects on the bias are entirely due to dependence of the collapse barrier on $q$.  

\end{document}